\documentclass[12pt]{iopart}
\usepackage{amsfonts}
\usepackage{graphicx}
\usepackage{epsfig}
\usepackage{color}

\def\beq{\begin{equation}}
\def\eeq{\end{equation}}
\def\beqa{\begin{eqnarray}}
\def\eeqa{\end{eqnarray}}

\def\hf{\textstyle\frac{1}{2}}

\def\quarter{\textstyle\frac{1}{4}}

\newcommand{\ket}[1]{\vert #1 \rangle}   

\newcommand{\bra}[1]{\langle #1 \vert}             

\newcommand{\mb}[1]{\mbox{\boldmath${#1}$}}
\newcommand{\smb}[1]{\mbox{\boldmath${\scriptstyle #1}$}}
\newcommand{\ssmb}[1]{\mbox{\boldmath${\scriptscriptstyle #1}$}}
\newcommand{\myfrac}[2]{\leavevmode\kern.1em
   \raise.5ex\hbox{\the\scriptfont0 #1}\kern-.1em
   /\kern-.15em\lower.25ex\hbox{\the\scriptfont0 #2}}

\newcounter{noproblem}
\newcounter{nosec}
\newcounter{noitem}

\begin{document}

\title[$\mathfrak{SU}(n)$ quasi-distributions]{General approach to $%
\mathfrak{SU}(n)$ quasi-distribution functions}
\author{Andrei~B Klimov$\,^1$ and Hubert~de~Guise$\,^2$}

\address{$\,^1$ Departamento de F\'{\i}sica, Universidad de
Guadalajara, 44420 Guadalajara, Jalisco, Mexico}

\address{$\,^2$ Department of Physics, Lakehead University, Thunder
Bay, Ontario P7B 5E1, Canada}

\begin{abstract}
We propose an operational form for the kernel of a
mapping between an operator acting in a Hilbert space of
a quantum system with $\mathfrak{SU}(n)$ symmetry group and its symbol in the
corresponding classical phase space. For symmetric irreps of
$\mathfrak{SU}(n)$, this mapping is bijective.  We briefly discuss complications
that will occur in the general case.
\end{abstract}


\maketitle


\section{Introduction}

There is renewed interest in physical systems involving higher symmetries,
motivated in part by recent experimental and theoretical results on
physical systems involving such higher symmetries.  Examples include
work on atomic and molecular systems \cite{molecule},
$n$-qubits in symmetric $SU(2^{n})$ state space, three--well Bose-Einstein condensates \cite{Viscondi}, general qudit systems, such as collections of distinguishable
$d$--level atoms.

These developments motivate a proper formulation of phase space methods adapted to higher symmetries. Phase space methods in quantum mechanics were pioneered by Wigner \cite{Wigner}, and
his work has been the seed for several hugely successful approaches having the
common objective of mapping quantum mechanical operators, defined in an
abstract Hilbert space, to complex-valued functions in a classical phase
space appropriate for the system under consideration \cite{reviews, Lee},
\cite{Perelomov}-\cite{Agarwal}. Several different methods for
Wigner-like mapping, applicable to a wide class of continuous (Lie type) and discrete
groups, have also appeared in the recent
literature \cite{Wolf} - \cite{mukunda2}.

Considerable insight into the possible mappings is provided by the axiomatic
Stratonovich-Weyl approach \cite{Moyal}-\cite{Beresin}, in which a
one-to-one correspondence between an operator $\hat{X}$ and its phase-space
symbol $W_{X}$ is established by restricting mappings to those having
\textquotedblleft reasonable\textquotedblright\ physical properties:

1) covariance, 2) hermiticity, 3) traciality and 4) normalization.

Within this framework, and using general ideas first advanced by
Berezin, an elegant form of the mapping kernel construction was proposed in
\cite{Brif}, where an explicit form for the $s$--ordered kernel
$\hat{w}^{(s)}(\Omega )$, automatically satisfying the conditions above, was also constructed.  The apparent simplicity of
the formulation of \cite{Brif} hides several technical complications. The
most unfortunate is that for higher groups, harmonic functions typically
depend on additional parameters: in $\mathfrak{SU}(3)$ for instance,
harmonic functions contain five parameters \cite{SU3harmonic} so one must be
eliminated ``by hand'' .

In this paper we propose an alternate but operational form of the $s$%
-ordered kernel, valid for $\mathfrak{SU}(n)$ irreducible representations
(irreps) of the type $(\lambda ,0,\ldots ,0)$.  The particular cases
of $\mathfrak{SU}(2)$ and $\mathfrak{SU}(3)$ are given in enough details for the
generalization to any symmetric representation of $\mathfrak{SU}(n)$.

\section{General setup and comments for the $(\protect\lambda,0,\ldots,0)$
irreps}

\subsection{Notation}

Throughout this paper we will use the shorthand $\lambda$ to mean the irrep $%
(\lambda,0,\ldots,0)$ of $\mathfrak{SU}(n)$. Exceptions to this abuse of
notation will be noted explicitly or will be clear from context.

Suppose the Hilbert space $\mathbb{H}$ for a quantum system carries a
unitary irreducible representation $\lambda $ of {a compact Lie group}
$\mathfrak{G}$,  {which we take henceforth to be $\mathfrak{SU}(n)$.} $\Lambda (g)$ is the matrix realization for irrep
$\lambda $ of the element $g\in \mathfrak{G}$, so that $\Lambda (g)$ acts by
linear transformations on $\mathbb{H}$.

For  {$\mathfrak{SU}(n)$} irreps of the type $(\lambda,0,\ldots ,0)$, states will be written $|\lambda ;
\mbox{\boldmath${\nu}$}\rangle $, where the $i$-th component of the weight $\mb{\nu}=(\nu_1,\ldots,\nu_{n-1})$
is the eigenvalue of the $i$'th Cartan element on the state:
\beq
\hat h_i\ket{\lambda; \mb{\nu}}=\nu_i\,\ket{\lambda; \mb{\nu}}\, .
\eeq

The highest weight $|\lambda ;\hbox{\rm h.w.}\rangle $ of irrep $\lambda$ is invariant under a
subgroup $\mathfrak{H}$ of $\mathfrak{G}$, and the phase space for the
corresponding classical system is isomorphic to the coset
$\mathfrak{M}=\mathfrak{G}/\mathfrak{H}$ \cite{Onofri}. $\mathfrak{G}$ acts on
$\mathfrak{M}$ by canonical transformations.

Operators acting on the Hilbert space carrying irrep $\lambda $ will
transform according to the irrep $\lambda \otimes \lambda ^{\ast }$, where
$\lambda ^{\ast }$ the irrep conjugate to $\lambda $.

The product $\lambda \otimes \lambda ^{\ast }$ is
reducible so, in addition to the labeling of states in irrep $\lambda$, we
must consider the labeling of tensors from a general irrep $\sigma =(\sigma _{1},\sigma
_{2},...,\sigma _{n-1})$.  
A tensor $\hat T^{\lambda}_{\sigma;\smb{\alpha}I_{\alpha}}$ is given explicitly by
\begin{equation}
\hat{T}_{\sigma ;\smb{ \gamma}I_{\gamma}}^{\lambda }
=\displaystyle\sum_{\alpha \beta }
\ket{\lambda ;\mb{\alpha}}\bra{\lambda ;\mb{\beta}^*}
\tilde C_{\lambda \smb{\alpha};\lambda ^{\ast }\, \smb{\beta}^{\ast }}^{\sigma \smb{ \gamma}I_{\gamma }}\,,
\label{Tdef}
\end{equation}
where $\lambda ^{\ast }$ the irrep conjugate to $\lambda $, $\beta ^{\ast }$
the weight conjugate to $\beta $, and $\sigma $ an irrep in the
decomposition of $\lambda \otimes \lambda ^{\ast }$.
Note that, for $\lambda
\equiv (\lambda ,0,\ldots ,0)$, $\sigma $ occurs at most once in
$\lambda \otimes \lambda ^{\ast }$.
The coefficients
$
\tilde C
_{\lambda \,\smb{\alpha};\lambda ^{\ast }\,\smb{\beta}^{\ast }}
^{\sigma \smb{\nu}I_{\nu }}
$ satisfy the orthogonality
relation
\begin{equation}
\sum_{\alpha \beta }\left( \tilde C_{\lambda \,\mbox{\boldmath${\scriptstyle
\alpha}$}\,;\lambda ^{\ast }\,\mbox{\boldmath${\scriptstyle \beta}$}^{\ast
}}^{\sigma ^{\prime }\,\mbox{\boldmath${\scriptstyle \nu}$}^{\prime }I_{\nu
^{\prime }}}\right) ^{\ast }\,\tilde C_{\lambda \,\mbox{\boldmath${%
\scriptstyle \alpha}$}\,;\lambda ^{\ast }\,\mbox{\boldmath${\scriptstyle
\beta}$}^{\ast }}^{\sigma \,\mbox{\boldmath${\scriptstyle \nu}$}I_{\nu
}}=\delta _{\sigma \sigma ^{\prime }}\delta _{\mbox{\boldmath${\scriptstyle
\nu}$}\mbox{\boldmath${\scriptstyle \nu}$}^{\prime }}\delta _{I_{\nu }I_{\nu
^{\prime }}}  \label{Tr_Ort}
\end{equation}%
and are elements of a unitary matrix.  The tensors satisfy
\beq
[\hat{h}_i,\hat T^{\lambda}_{\sigma; \smb{\alpha}I_{\alpha}}]=\alpha_i\,\hat T^{\lambda}_{\sigma; \smb{\alpha}I_{\alpha}}\, .
\eeq
In general, some weights in $\sigma$ will occur multiple times;
$I_{\alpha }$ distinguishes between multiple occurrences of the
weight $\mbox{\boldmath${\alpha}$}$.  
Eqn.(\ref{Tr_Ort}) implies that the tensors are trace orthogonal over $\sigma ,%
\mbox{\boldmath${\alpha}$}$ and $I_{\alpha }$,

\subsection{Construction of the kernel}

According to the Stratonovich--Weyl method, we construct a Hermitian kernel
$\hat{w}^{(s)}(\Omega )$, $\Omega \in \mathfrak{M}$, that implements a
mapping between operators $\hat{X}$ acting in $\mathbb{H}$ and symbols $%
W_{X}\in \mathfrak{M}$ via
\begin{equation}
\begin{array}{rcccl}
\hat{X} & \stackrel{{\hat{w}}^{(s)}(\Omega )}{\longrightarrow } &
W_{X}^{(s)}(\Omega ) & = & \hbox{\rm Tr}\left( \hat{X}\hat{w}^{(s)}(\Omega
)\right) \,, \\
W_{X}^{(s)}(\Omega ) & \stackrel{\hat{w}^{(-s)}(\Omega )}{\longrightarrow } &
\hat{X} & = & \displaystyle\int d\Omega \,\hat{w}^{(-s)}(\Omega
)W_{X}^{(s)}(\Omega )\,.%
\end{array}%
\end{equation}
Here, $s$ is a (continuous) ordering parameter, which takes values $s=-1,0$
and $1$ for the normal, symmetric and anti-normal ordering of
operators respectively.

The kernel $\hat{w}^{(s)}(\Omega )$ can always be written in the form,
\begin{equation}
\hat{w}^{(s)}(\Omega )=\Lambda(\Omega )\hat{P}^{(s)}\Lambda^{\dag } (\Omega
),\quad \Omega \in \mathfrak{M}\, ,\label{wingeneralform}
\end{equation}%
The essential information about the mapping is contained in the operator
$\hat{P}^{(s)}$, which we write in the integral form
\begin{equation}
\hat{P}^{(s)}=\displaystyle\int d\omega \,\e^{i\omega \hat{h}%
_{1}}\,f^{(s)}(\omega )\,,  \label{operatorP}
\end{equation}%
with $f^{(s)}(\omega )$ a scalar function to be determined, and $\hat{h}_{1}$
the $\mathfrak{H}$-invariant Cartan element:
\begin{equation}
\Lambda (\bar{g})\,\hat{h}_{1}\,\Lambda^{\dagger } (\bar{g})=
\hat{h}_{1}\,,\qquad \bar{g}\in \mathfrak{H}\, .  \label{diagonalpiece}
\end{equation}

This form for $\hat{P}^{(s)}$ guarantees the covariance of
$\hat{w}^{(s)}(\Omega )$ under transformations from the coset
$\mathfrak{G}/\mathfrak{H}$:
\begin{equation}
\Lambda(\tilde{\Omega})\hat{w}^{(s)}(\Omega )\Lambda ^{\dagger }(\tilde{\Omega})
=\hat{w}^{(s)}(\tilde\Omega \Omega)\,,\quad \tilde{\Omega}\in
\mathfrak{G}/\mathfrak{H}\,.
\end{equation}

 {To determine $f^{(s)}$ we proceed as follows.}
For symmetric irreps of $\mathfrak{SU}(n)$, the highest weight vector is
$\mathfrak{U}(n-1)$ invariant. This  {forces} $\hat{h}_{1}$ (in
the defining $n\times n$ irrep) to be
\begin{equation}
\hat{h}_{1}\equiv \hbox{\rm diag}(n-1,-1,\ldots ,-1)\,.
\end{equation}

Using the tensors defined in Eqn.(\ref{Tdef}), one  {then} expands $\e^{i \omega \hat h_1}$ as
\begin{equation}
\e^{i \omega \hat h_1}=\displaystyle\sum_{\sigma}\tilde\chi^\lambda_{%
\sigma}(\omega)\, \hat T^\lambda_{\sigma;\mbox{\boldmath${\scriptstyle 0}$}%
0}\, ,  \qquad  {\tilde \chi^\lambda_\sigma(\omega)=\hbox{\rm Tr}\left[\e^{i\omega\hat h_1}\,
\left(\hat T^\lambda_{\sigma;\mbox{\boldmath${\scriptstyle 0}$}0}\right)^\dagger\right]} \label{expansionofexp}
\end{equation}
of zero--weight tensors $\hat T^\lambda_{\sigma;\mbox{\boldmath${\scriptstyle 0}$}0}$ that must carry the trivial $I_0\equiv 0$ irrep of $ {\mathfrak{H}}=\mathfrak{U}(n-1)$.

For later convenience, we note that one can also expand $\Lambda (\Omega )%
\hat{T}_{\sigma ;\mbox{\boldmath${\scriptstyle 0}$}0}^{\lambda }\Lambda
^{\dagger }(\Omega )$ to get
\begin{eqnarray}
\hat{w}^{(s)}(\Omega ) &=&\sum_{\sigma }F_{\sigma }^{(s)}\sum_{%
\mbox{\boldmath${\scriptstyle \beta}$}I_{\beta }}\,D_{\mbox{\boldmath${%
\scriptstyle \beta}$}I_{\beta };\mbox{\boldmath${\scriptstyle 0}$}0}^{\sigma
}(\Omega )\hat{T}_{\sigma ;\mbox{\boldmath${\scriptstyle \beta}$}I_{\beta
}}^{\lambda }\,,  \label{wexpandedinD} \\
F_{\sigma }^{(s)} &=&\displaystyle\int \,d\omega \,\tilde{\chi}_{\sigma
}^{\lambda }(\omega )\,f^{(s)}(\omega )\,  \label{Fexpandedinf}
\end{eqnarray}%
with
\begin{equation}
D_{\smb{\beta}I_{\beta };\smb{0}0}^{\sigma }(\Omega )\equiv \langle \sigma ;%
\mbox{\boldmath${\beta}$}I_{\beta }|\Lambda (\Omega )|\sigma ;%
\mbox{\boldmath${0}$}0\rangle ,
\end{equation}%
an $\mathfrak{SU}(n)$ group function for the irrep $\sigma $ \cite{su3Wigner}.

To accommodate the requirement that hermitian operators are mapped into real
functions, we  {demand that } $\hat{w}^{(s)}(\Omega )$ be hermitian, which in
turn implies
\begin{equation}
\hat{w}^{(s)}(\Omega )=\left( \hat{w}^{(s)}(\Omega )\right) ^{\dagger }\
\Rightarrow \ \left( f^{(s)}(\omega )\right) ^{\ast }=f^{(s)}(-\omega )\,.
\label{hermiticityofw}
\end{equation}

To guarantee the invertibility of the map, we impose a \emph{traciality}
condition on $\hat w^{(s)}(\Omega)$, expressed by
\begin{equation}
\hbox{\rm Tr}\left(\hat w^{(s)}(\Omega^{\prime})^\dagger\,\hat
w^{(-s)}(\Omega)\right)=N^\lambda\Delta(\Omega^{\prime},\Omega)\, ,
\label{tracialitycondition}
\end{equation}
where $N^\lambda$ is the proportionality constant, and $\Delta(\Omega^{%
\prime},\Omega)$ is the self-reproducing kernel
\begin{equation}
\displaystyle\int d\Omega\,
z(\Omega)\,\Delta(\Omega^{\prime},\Omega)=z(\Omega^{\prime})\, ,\qquad
\Omega,\Omega^{\prime}\in \mathfrak{M}\, ,
\end{equation}
valid for any function $z$ on $\mathfrak{M}$.

The traciality condition can be expanded to produce
\begin{equation}
N^{\lambda }\Delta (\Omega ^{\prime },\Omega )=\displaystyle\sum_{\sigma
}\left( F_{\sigma }^{(s)}\right) ^{\ast }\,F_{\sigma }^{(-s)}\,\sum_{%
\mbox{\boldmath${\scriptstyle \alpha}$}I_{\alpha }}\left( D_{%
\mbox{\boldmath${\scriptstyle \alpha}$}I_{\alpha },\mbox{\boldmath${%
\scriptstyle 0}$}0}^{\sigma }(\Omega ^{\prime })\right) ^{\ast }\,D_{%
\mbox{\boldmath${\scriptstyle \alpha}$}I_{\alpha },\mbox{\boldmath${%
\scriptstyle 0}$}0}^{\sigma }(\Omega )\,,  \label{tracialityinD}
\end{equation}%
where the orthogonality condition (\ref{Tr_Ort}) has been used. This is a
restriction on $F_{\sigma }^{(s)}$ and thus on the functions $f^{(s)}(\omega)$,
which expand $F_{\sigma }^{(s)}$ via Eqn.(\ref{Fexpandedinf}).

Since the functions \{$D_{\smb{\beta}I_{\beta };\smb{0}0}^{\sigma }(\Omega )$\} are complete and orthogonal under integration over
the coset, the reproducing kernel can be  {written} as
\begin{equation}
\Delta(\Omega^{\prime},\Omega)=\sum_{\sigma \mbox{\boldmath${\scriptstyle
\alpha}$}I_\alpha} \left( \frac{\hbox{\rm vol}(\mathfrak{M})}{\hbox{\rm dim}%
(\sigma)}\right) \left(D^{\sigma}_{\mbox{\boldmath${\scriptstyle \alpha}$}%
I_{\alpha},\mbox{\boldmath${\scriptstyle 0}$} 0}(\Omega^{\prime})\right)^*\,
D^\sigma_{\mbox{\boldmath${\scriptstyle \alpha}$}I_{\alpha},%
\mbox{\boldmath${\scriptstyle 0}$} 0}(\Omega)\, ,  \label{DeltainD}
\end{equation}
where $\hbox{\rm vol}(\mathfrak{M})$ is related to the volume of $\mathfrak{M%
}$ calculated using the invariant measure $d\Omega$ and $\hbox{\rm dim}%
(\sigma)$ is the dimension of irrep $\sigma$.

Combining Eqns.(\ref{DeltainD}) and (\ref{tracialityinD}), and observing
that the irrep $\sigma $ occurs at most once in $\lambda \otimes \lambda
^{\ast }$, we find a cross--condition on dual $F_{\sigma }^{(s)}$'s:
\begin{equation}
N^{\lambda }\,\frac{\hbox{\rm vol}(\mathfrak{M})}{\hbox{\rm dim}(\sigma )}%
=\left( F_{\sigma }^{(s)}\right) ^{\ast }\,F_{\sigma }^{(-s)}
\label{crosscondition}
\end{equation}

The normalization property is simply
\begin{equation}
\hbox{\rm Tr}\left( \hat{w}^{(s)}(\Omega )\right) =1=F_{0}^{(s)}\,\sqrt{%
\hbox{\rm dim}(\lambda )},  \label{normalizationproperty}
\end{equation}%
as all tensors but the $\sigma =0$ (scalar)
representation in $\lambda \otimes \lambda ^{\ast }$ are traceless. We can
use this in Eqn.(\ref{crosscondition}) to obtain
\begin{equation}
N^{\lambda }=\frac{1}{\hbox{\rm vol}(\mathfrak{M})\hbox{\rm dim}(\lambda )}%
\,.  \label{normalizationN}
\end{equation}%
and feed this back in Eqn.(\ref{crosscondition}) to obtain more generally
\begin{equation}
\left( F_{\sigma }^{(s)}\right) ^{\ast }\,F_{\sigma }^{(-s)}=\frac{1}{%
\hbox{\rm dim}(\lambda )\hbox{\rm dim}(\sigma )}\,.  \label{FFstar}
\end{equation}

Next, we make the important observation that, for $s=-1$, the usual $Q$%
-function kernel should be recovered:
\begin{equation}
\hat{w}^{(-1)}(\Omega )=\Lambda(\Omega )|\lambda ;\hbox{\rm h.w.}%
\rangle \langle \lambda ;\hbox{\rm h.w.}|\Lambda^{\dagger } (\Omega )\,.
\end{equation}%
This implies the ``boundary condition''
\begin{equation}
\hat P^{(-1)}=|\lambda ;\hbox{\rm h.w.}\rangle \langle \lambda ;\hbox{\rm h.w.}|=%
\displaystyle\int d\omega \,\e^{i\omega \hat{h}_{1}}\,f^{(-1)}(\omega
)=\sum_{\sigma }\,F_{\sigma }^{(-1)}\,\hat{T}_{\sigma ;\mbox{\boldmath${%
\scriptstyle 0}$}0}^{\lambda }\,,
\end{equation}%
from which
\begin{equation}
F_{\sigma }^{(-1)}=\tilde{C}_{\lambda \,\hbox{\scriptsize h.w.};\lambda
^{\ast }\,\hbox{\scriptsize h.w.}^{\ast }}^{\sigma \mbox{\boldmath${%
\scriptstyle 0}$}0}=\langle \lambda ;\hbox{\rm h.w.}|\hat{T}_{\sigma ;%
\mbox{\boldmath${\scriptstyle 0}$}0}^{\lambda }|\lambda ;\hbox{\rm h.w.}%
\rangle \,.  \label{bigFminus1}
\end{equation}%
Combining Eqns.(\ref{FFstar}) and (\ref{bigFminus1})  {yields}
\begin{equation}
F_{\sigma }^{(1)}=\frac{1}{\hbox{\rm dim}(\lambda )\hbox{\rm dim}(\sigma
)\left( \tilde{C}_{\lambda \,\hbox{\scriptsize h.w.};\lambda ^{\ast }\,%
\hbox{\scriptsize h.w.}^{\ast }}^{\sigma \mbox{\boldmath${\scriptstyle 0}$}%
0}\right) ^{\ast }}\,.
\end{equation}%
We can interpolate to arbitrary $s$ by defining
\begin{equation}
F_{\sigma }^{(s)}=\frac{\left( \tilde{C}_{\lambda \,\hbox{\scriptsize h.w.}%
;\lambda ^{\ast }\,\hbox{\scriptsize h.w.}}^{\sigma \mbox{\boldmath${%
\scriptstyle 0}$}0}\right) ^{-s}}{\left[ \hbox{\rm dim}(\lambda )\hbox{\rm
dim}(\sigma )\right] ^{(s+1)/2}}\,.  \label{Fforanys}
\end{equation}

We are now in a position to determine $f^{(s)}$. If we suppose
\begin{equation}
f^{(-1)}(\omega )=\sum_{\sigma }c_{\sigma }^{(-1)}\,\left( \tilde{\chi}%
_{\sigma }^{\lambda }(\omega )\right) ^{\ast }
\end{equation}%
then
\begin{equation}
\tilde{C}_{\lambda \,\hbox{\scriptsize h.w.};\lambda ^{\ast }\,%
\hbox{\scriptsize h.w.}^{\ast }}^{\sigma \mbox{\boldmath${\scriptstyle 0}$}%
0}=\sum_{\sigma ^{\prime }}g_{\sigma \sigma ^{\prime }}\,c_{\sigma ^{\prime
}}^{(-1)}\,,\qquad g_{\sigma \sigma ^{\prime }}=\displaystyle\int d\omega \,%
\tilde{\chi}_{\sigma }^{\lambda }(\omega )\,\left( \tilde{\chi}_{\sigma
^{\prime }}^{\lambda }(\omega )\right) ^{\ast }.
\end{equation}%
Since
\begin{equation}
\int d\omega \,\left( f^{(-1)}(\omega )\right) ^{\ast }\,f^{(-1)}(\omega
)=\sum_{\sigma ,\sigma ^{\prime }}c_{\sigma }^{(-1)}\,g_{\sigma \sigma
^{\prime }}\left( c_{\sigma ^{\prime }}^{(-1)}\right) ^{\ast }>0\,.
\end{equation}%
the overlap matrix $g_{\sigma \sigma ^{\prime }}$ is necessarily invertible.
Writing $g^{\mu \nu }$ as $(g)_{\mu \nu }^{-1}$
we can solve for $c_{\sigma'}^{(-1)}$ and,
extending the expansion of $f$ to any $s$, obtain
\begin{equation}
f^{(s)}(\omega)=\sum_{\sigma}c^{(s)}_{\sigma}\,\left(\tilde\chi^{\lambda}_{%
\sigma}(\omega)\right)^* \, ,\quad
c^{(s)}_{\sigma}=\sum_{\mu}g^{\sigma\mu}\,F^{(s)}_\mu\, .  \label{generalcs}
\end{equation}

 {We can combine all the relevant equations and obtain the following simplified form for $\hat P^{(s)}$:
\begin{eqnarray}
\hat P^{(s)} &=&\int d\omega\, \e^{i\omega \hat h_{1}}\,\left( \sum_{\sigma '}\,c_{\sigma '}^{(s)}\left( \tilde{\chi }_{\sigma'}^{\lambda }(\omega )\right)
^{* }\right)  \\
&=&\sum_{\sigma \sigma '}\hat T_{\sigma ;\mbox{\boldmath${\scriptstyle 0}$}0}^{\lambda }
\left( \int d\omega\, \tilde{\chi }_{\sigma }^{\lambda }(\omega )
\left( \tilde{\chi }_{\sigma '}^{\lambda }(\omega )\right) ^{*}\right)
c_{\sigma ' }^{(s)} \\
&=&\sum_{\sigma \sigma '}\hat T_{\sigma ;\mbox{\boldmath${\scriptstyle 0}$}0}^{\lambda }
\,g_{\sigma \sigma '}\sum_{\mu }g^{\sigma ' \mu }F_{\mu }^{(s)} \\
&=&\sum_{\sigma \mu }\hat T_{\sigma ;\mbox{\boldmath${\scriptstyle 0}$}0}^{\lambda }
\left( \sum_{\sigma ' }\, g_{\sigma \sigma ' }g^{\sigma ' \mu }\right) F_{\mu }^{(s)}
\\
&=&\sum_{\sigma }\hat T_{\sigma ;\mbox{\boldmath${\scriptstyle 0}$}0}^{\lambda }\, \frac{\left( \tilde{C}_{\lambda \,\hbox{\scriptsize h.w.};\lambda ^{\ast }\,%
\hbox{\scriptsize h.w.}^{\ast }}^{\sigma \mbox{\boldmath${\scriptstyle 0}$}%
0}\right) ^{-s}}{\left[ \dim
\left( \lambda \right) \dim \left( \sigma \right) \right] ^{(s+1)/2}}
\end{eqnarray}
}

\section{The case of $\mathfrak{SU}(2)$}

Basis states for the irrep $j$ are taken as usual to be $\{|jm\rangle \}$,
such that
\begin{equation}
\hat{S}_{z}|jm\rangle =m|jm\rangle \,.
\end{equation}%
The highest weight of irrep $j$ is $|jj\rangle $ and invariant (up to a
phase) under transformations of the form $\e^{i\omega \hat{S}_{z}}$. Thus, $%
\mathfrak{H}$ is the $\mathfrak{U}(1)$  {subgroup} generated by $\e^{i\omega \hat{S}_{z}}
$, and $\mathfrak{M}=\mathfrak{SU}(2)/\mathfrak{U}(1) {\sim S^2}$. Coset
representatives are taken to be
\begin{equation}
\Lambda (\Omega )=R_{z}(\alpha )R_{y}(\beta )\equiv \e^{i\alpha \hat{S}%
_{z}}\,\e^{i\beta \hat{S}_{y}}\,.
\end{equation}%
and $\hat{P}^{(s)}$ is written simply as the integral
\begin{equation}
\hat{P}^{(s)}=\displaystyle\int d\omega \,\e^{i\omega \hat{S}%
_{z}}\,f^{(s)}(\omega )\,.
\end{equation}%
The trace--orthogonal $\mathfrak{SU}(2)$ tensor operators are
\begin{equation}
\hat{T}_{LM}^{j}=\sum_{mm^{\prime }}|jm\rangle \langle jm^{\prime
}|\,C_{j\,m\,;j\,-m^{\prime }}^{L\,M}\,(-1)^{j-m}\,,
\end{equation}%
where $C_{j\,m\,;j\,-m^{\prime }}^{L\,M}$ is the usual Clebsch-Gordan (CG)
coefficient for $\mathfrak{SU}(2)$. We expand
\begin{equation}
\e^{i\omega \hat{S}_{z}}=\displaystyle\sum_{L}\tilde{\chi}_{L}^{j}(\omega )\,%
\hat{T}_{L0}^{j}\,,\quad \tilde{\chi}_{L}^{j}(\omega )=\displaystyle%
\sum_{m}\left( C_{j\,m\,;j\,-m}^{L\,0}\,(-1)^{j-m}\right) \,\e^{i\omega m}\,,
\label{expansionofexpSz}
\end{equation}%
in terms of the zero--weight tensors $\hat{T}_{L0}^{j}$. With this
Eqns. (\ref{DeltainD}) and (\ref{normalizationN}) specialize
to:
\begin{eqnarray}
\Delta (\Omega ^{\prime },\Omega ) &=&\sum_{L}\left( \frac{4\pi }{2L+1}%
\right) \left( D_{M0}^{L}(\Omega ^{\prime })\right) ^{\ast
}\,D_{M0}^{L}(\Omega )\,, \\
N^{j} &=&\frac{1}{4\pi (2j+1)}\,,
\end{eqnarray}%
with $D_{M0}^{L}$ the usual Wigner $D$-function is proportional to $%
Y_{LM}^{\ast }(\beta ,\alpha )$.

For $\mathfrak{SU}(2)$, the generalized characters $\tilde{\chi}%
_{L}^{j}(\omega )$ satisfy the orthogonality relation $g_{L^{\prime }L}=2\pi
\,\delta _{L^{\prime }L}$. leading to a closed form solution is possible for
any $j$ since, by Eqn.(\ref{generalcs}), $c_{L}^{(s)}=\left( 2\pi \right)
^{-1}\,F_{L}^{(s)}\,,$ and
\begin{equation}
F_{L}^{(s)}=\frac{\left(C^{L0}_{jm,j-m}\right)^{-s}}{\left[
(2j+1)(2L+1)\right] ^{(s+1)/2}}\,,
\end{equation}%
as per Eqn.(\ref{Fforanys}). Hence, the final expression for $f^{(s)}$ is
\begin{equation}
f^{(s)}(\omega)=\frac{1}{2\pi }\sum_{L}\,F_{L}^{(s)}\,\left( \tilde{\chi}%
_{L}^{j}(\omega )\right) ^{\ast }\,.
\end{equation}

\section{The case of $\mathfrak{SU}(3)$ irreps of the type $(\protect\lambda%
,0)$.}

We label states of the $\mathfrak{SU}(3)$ irrep
 $(\lambda,0 )$ by $|\lambda; \mb{\nu}\rangle$.
The weight $\mb{\nu}\equiv (\nu _{1}-\nu _{2},\nu -\nu _{3})$ of the state
is extracted from a triple $[\nu _{1},\nu_{2},\nu _{3}]$ on non-negative integers constrained
by $\nu _{1}+\nu _{2}+\nu _{3}=\lambda $.
The irrep $(\lambda,0)$ does not have weight multiplicities, \emph{i.e.}
each weight $\mbox{\boldmath${\nu}$}$ occurs at most once. In $(\lambda,0)$, the multiplicity label $I\equiv I_{23}$
of \cite{su3Wigner} is
redundant but given by $I_{23}=\textstyle{\frac{1}{2}}
(\nu_2+\nu_3)$.

We denote by $(\lambda,0)^* \equiv (0,\lambda)$ the irrep conjugate to $%
(\lambda,0)$ and recall
\begin{equation}
(\lambda,0)\otimes (\lambda,0)^* =(0,0)\oplus (1,1)\oplus \ldots \oplus
(\lambda,\lambda)\, =\sum_{\sigma=0}^\lambda (\sigma,\sigma).
\end{equation}

Following the template of \cite{su3Wigner}, one shows
that $\mathfrak{SU}(3)$ transformations can be factorized as
\begin{eqnarray}
&&R(\alpha _{1},\beta _{1},\alpha _{2},\beta _{2},\alpha _{3},\beta
_{3},\gamma _{1},\gamma _{2})  \nonumber \\
&&\quad =R_{23}(\alpha _{1},\beta _{1},-\alpha _{1})\cdot R_{12}(\alpha
_{2},\beta _{2},-\alpha _{2})\,  \nonumber \\
&&\qquad \cdot R_{23}(\alpha _{3},\beta _{3},-\alpha _{3})\cdot \e^{-i\gamma
_{1}(C_{11}-C_{22})}\,\e^{-i\gamma _{2}(C_{22}-C_{33})}\,.
\end{eqnarray}

The highest weight is $\vert (\lambda,0)(\lambda,0) \rangle$. It is an $I_{23}=0$ state invariant (up to a phase) under transformations of the subgroup $\mathfrak{H}=%
\mathfrak{U}_{23}(2)$ generated by
$R_{23}(\alpha_3,\beta_3,-\alpha_3)\,\e^{-i\gamma_1(C_{11}-C_{22})}\,\e%
^{-i\gamma_2(C_{22}-C_{33})}$. In the fundamental representation $(1,0)$, the
matrices of $\mathfrak{U}_{23}(2)$ are of the form
\begin{equation}
\mathfrak{U}_{23}(2)\sim \left(%
\begin{array}{ccc}
* & 0 & 0 \\
0 & * & * \\
0 & * & *%
\end{array}%
\right)
\end{equation}
with $*$ denoting a non-zero entry. Elements $\Omega$ in the coset $%
\mathfrak{SU}(3)/\mathfrak{U}_{23}(2)$ correspond to transformations of the
form
\begin{equation}
\Lambda(\Omega)=
R_{23}(\alpha_1,\beta_1,-\alpha_1)R_{12}(\alpha_2,\beta_2,-\alpha_2)\, ,
\end{equation}
with parameter range $0\le \alpha_1,\alpha_2\le 2\pi$, $0\le\beta_1,\beta_2\le \pi$.
The measure on the coset is
\begin{equation}
d\Omega=\sin(\beta_1)\cos(\textstyle{\frac{1}{2}}\beta_2)\sin^{3}(\textstyle{%
\frac{1}{2}}\beta_2)\,d\alpha_1\,d\beta_1\,d\alpha_2\,d\beta_2
\end{equation}
and the coset volume is $4\pi^2$.

The operator $\hat P^{(s)}$ of Eq.(\ref{operatorP}) is now
\begin{equation}
\hat P^{(s)}=\displaystyle\int d\omega\,\e^{i\omega\hat
h_1}\,f^{(s)}(\omega)\ .
\end{equation}
In the fundamental, $(1,0)$ irrep, we have
\begin{equation}
\e^{i\omega\hat h_1}=\hbox{\rm
diag}\left(\e^{2i\omega},\e^{-i\omega},\e^{-i\omega}\right)\, .
\end{equation}

The expansion of $\e^{i\omega\hat h_1}$ will be a sum of diagonal terms of
the form
\beq
\e^{i\omega\hat
h_1}=\sum_{\sigma=0}^{\lambda}\tilde\chi^{(\lambda,0)}_{(\sigma,\sigma)}(%
\omega)\hat T^{(\lambda,0)}_{(\sigma,\sigma),\mbox{\boldmath${\scriptstyle
0}$}0}\, ,  \label{chidefinition}
\eeq
with $\mb{0}$ the zero weight; there can be more than one copy of this weight
in $(\sigma,\sigma)$ but only $I_{23}=0$ appear so  {as} to ensure $\mathfrak{U}_{23}$--invariance of
$\e^{i\omega\hat h_1}$.
The coset functions are
\begin{equation}
D^{(\sigma,\sigma)}_{\mbox{\boldmath${\scriptstyle \nu}$}I;%
\mbox{\boldmath${\scriptstyle 0}$}0}(\Omega)\equiv \langle (\sigma,\sigma)%
\mbox{\boldmath${\nu}$}I \vert R_{23}(\alpha_1,\beta_1,-\alpha_1)
R_{12}(\alpha_2,\beta_2,-\alpha_2)\vert (\sigma,\sigma)\mbox{\boldmath${0}$}%
0 \rangle
\end{equation}
with $\Omega$ now in $\mathfrak{M}=\mathfrak{SU}(3)/\mathfrak{U}(2)$.

Since $\hbox{\rm dim}(\lambda,0)=\textstyle{\frac{1}{2}}\,(\lambda+1)(%
\lambda+2)$ and $\hbox{\rm dim}(\sigma,\sigma)=(\sigma+1)^3$, Eqns. (\ref{DeltainD}%
) and (\ref{normalizationN}) now specialize, for $\mathfrak{SU}(3)$ irreps
of the type $(\lambda,0)$ to:
\begin{eqnarray}
\Delta(\Omega^{\prime},\Omega)&=&\displaystyle\sum_{\mbox{\boldmath${%
\scriptstyle \nu}$}I\sigma}\frac{4\pi^2}{(\sigma+1)^3} \left(D^{(\sigma,%
\sigma)}_{\mbox{\boldmath${\scriptstyle \nu}$}I;\mbox{\boldmath${%
\scriptstyle 0}$}0}(\Omega^{\prime})\right)^* D^{(\sigma,\sigma)}_{%
\mbox{\boldmath${\scriptstyle \nu}$}I;\mbox{\boldmath${\scriptstyle 0}$}%
0}(\Omega) \, , \\
N^{\lambda}&=&\frac{1}{4\pi^2\, \textstyle{\frac{1}{2}}(\lambda+1)(%
\lambda+2)}\, ,
\end{eqnarray}

In the case of $\mathfrak{SU}(3)$, the various $\mathfrak{U}_{23}(2)$
subspaces will typically contain more than one weight. This in turn leads to
non-trivial overlap relations $g_{\sigma \sigma ^{\prime }}$ between various
$\tilde{\chi}_{(\sigma ,\sigma )}^{(\lambda ,0)}(\omega )$'s, defined by Eq.(%
\ref{chidefinition}).

\subsection{The case of $(1,0)$}

Direct calculation shows the zero-weight tensors related
to uncoupled projectors by
\begin{equation}
\left( {\renewcommand{\arraystretch}{1.5}
\begin{array}{c}
\hat T_{(0,0);\smb{0}0}^{(1,0)} \\
\hat T_{(1,1);\smb{0}1}^{(1,0)} \\
\hat T_{(1,1);\smb{0}0}^{(1,0)}%
\end{array}%
}\right) =\left( {\renewcommand{\arraystretch}{1.5}\renewcommand{\arraycolsep}{0.5pt}
\begin{array}{ccc}
\frac{1}{\sqrt{3}} & \frac{1}{\sqrt{3}} & \frac{1}{\sqrt{3}} \\
0 & -\frac{1}{\sqrt{2}} & \frac{1}{\sqrt{2}} \\
\sqrt{\frac{2}{3}} & -\frac{1}{\sqrt{6}} & -\frac{1}{\sqrt{6}}%
\end{array}%
}\right) \left( {\renewcommand{\arraystretch}{1.5}%
\begin{array}{c}
\left\vert (1,0)(10)\right\rangle \left\langle (1,0)(10)\right\vert  \\
\left\vert (1,0)(-11)\right\rangle \left\langle (1,0)(-11)\right\vert  \\
\left\vert (1,0)(0-1)\right\rangle \left\langle (1,0)(0-1)\right\vert
\end{array}%
}\right) ,
\end{equation}
with $\mb{0}=(00)$ denoting the zero weight.
The highest weight projector is given by $|(1,0)(10)\rangle \langle (1,0)(10)|$, and
Eqn.(\ref{Fforanys}) gives
\begin{equation}
F_{0}^{(s)}=\frac{\left( \sqrt{\leavevmode\kern.1em\raise.5ex%
\hbox{\the\scriptfont0 1}\kern-.1em/\kern-.15em\lower.25ex%
\hbox{\the\scriptfont0 3}}\right) ^{-s}}{3^{(s+1)/2}}\,,\qquad F_{1}^{(s)}=%
\frac{\left( \sqrt{\leavevmode\kern.1em\raise.5ex\hbox{\the\scriptfont0 2}%
\kern-.1em/\kern-.15em\lower.25ex\hbox{\the\scriptfont0 3}}\right) ^{-s}}{%
(3\cdot 8)^{(s+1)/2}}
\end{equation}%
Moreover, one rapidly verifies that for $(1,0)$
\begin{eqnarray}
\hbox{\rm e}^{i\omega \hat{h}_{1}}
&=&\hbox{\rm e}^{2i\omega }\left\vert (1,0)(10)\right\rangle \left\langle (1,0)(10)\right\vert   \nonumber \\
&+&\hbox{\rm e}^{-i\omega }\Bigl[|(1,0)(-1-1)\rangle \langle (1,0)(-1-1)|
\nonumber \\
&&\qquad \quad +|(1,0)(0-1)\rangle \langle (1,0)(0-1)|\Bigr] \\
&=&
\textstyle\frac{1}{\sqrt{3}}\left(2\hbox{\rm e}^{-i\omega }+\hbox{\rm e}^{2i\omega }\right) T_{(0,0);\smb{0}0}^{(1,0)}
+\textstyle\sqrt{\frac{2}{3}}\,
\left( -\hbox{\rm e}^{-i\omega }+\hbox{\rm e}^{2i\omega }\right)
T_{(1,1);\smb{0}0}^{(1,0)},
\end{eqnarray}
so that
\begin{equation}
\tilde{\chi}_{(0,0)}^{(1,0)}(\omega _{1})=\textstyle\frac{1}{\sqrt{3}}\left(2\hbox{\rm e}^{-i\omega
_{1}}+\hbox{\rm e}^{2i\omega _{1}}\right)\,,\quad \chi
_{(1,1)}^{(1,0)}(\omega _{1})=\textstyle\sqrt{\frac{2}{3}}\left( -\hbox{\rm e}%
^{-i\omega _{1}}+\hbox{\rm e}^{2i\omega _{1}}\right) \,.
\end{equation}%
With these we can form the overlap matrix
\begin{equation}
g_{\sigma \sigma ^{\prime }}=\left(
\begin{array}{cc}
\frac{10\pi }{3} & -\frac{2\sqrt{2}\pi }{3} \\
-\frac{2\sqrt{2}\pi }{3} & \frac{8\pi }{3}%
\end{array}%
\right) \,,
\end{equation}%
From the inverse $g^{\sigma\sigma'}$, the final form of the expansion coefficients $c_{\sigma}^{(s)}$ is thus
\begin{equation}
c_{0}^{(s)}=\frac{8+4^{-s}}{24\sqrt{3}\pi }\,,\qquad c_{1}^{(s)}=\frac{%
4+5\cdot 4^{-s}}{24\sqrt{6}\pi }\,.
\end{equation}%
For the special cases $s=-1,0,1$ we obtain
\begin{equation}
f^{(-1)}(\omega )=\frac{\e^{-2i\omega }}{2\pi }\,,\;f^{(0)}(\omega )=\frac{\e^{i\omega }+2\e^{-2i\omega }}{8\pi }\,,\;f^{(1)}(\omega )=\frac{5\e%
^{i\omega }+6\e^{-2i\omega }}{32\pi }\,.
\end{equation}

The operator $\hat P^{(0)}$ has the form diag$(\hf,\quarter,\quarter)$.
This case has also been investigated from a different approach in \cite{AlfredoLuis}.

\subsection{The case of $(2,0)$}

The transformation between zero-weight tensors and diagonal projectors is
given by
\begin{eqnarray}
\left(
{\renewcommand{\arraystretch}{1.25}
\begin{array}{c}
\hat{T}_{(0,0);\smb{0}0}^{(2,0)} \\
\hat{T}_{(1,1);\smb{0}1}^{(2,0)} \\
\hat{T}_{(1,1);\smb{0}0}^{(2,0)} \\
\hat{T}_{(2,2);\smb{0}2}^{(2,0)} \\
\hat{T}_{(2,2);\smb{0}1}^{(2,0)} \\
\hat{T}_{(2,2);\smb{0}0}^{(2,0)}%
\end{array}}
\right)  &=&U\left(
{\renewcommand{\arraystretch}{1.25}
\begin{array}{c}
\left\vert (2,0)(20)\right\rangle \left\langle (2,0)(20)\right\vert  \\
\left\vert (2,0)(01)\right\rangle \left\langle (2,0)(01)\right\vert  \\
\left\vert (2,0)(1-1)\right\rangle \left\langle (2,0)(1-1)\right\vert  \\
\left\vert (2,0)(-22)\right\rangle \left\langle (2,0)(-22)\right\vert  \\
\left\vert (2,0)(-10)\right\rangle \left\langle (2,0)(-10)\right\vert  \\
\left\vert (2,0)(0-2)\right\rangle \left\langle ((2,0)(0-2)\right\vert
\end{array}}
\right) \,,  \nonumber \\
U &=&\left(
{\renewcommand{\arraystretch}{1.5}\renewcommand{\arraycolsep}{0.75pt}
\begin{array}{cccccc}
\frac{1}{\sqrt{6}} & \frac{1}{\sqrt{6}} & \frac{1}{\sqrt{6}} & \frac{1}{%
\sqrt{6}} & \frac{1}{\sqrt{6}} & \frac{1}{\sqrt{6}} \\
0 & -\frac{1}{\sqrt{10}} & \frac{1}{\sqrt{10}} & -\sqrt{\frac{2}{5}} & 0 &
\sqrt{\frac{2}{5}} \\
2\sqrt{\frac{2}{15}} & \frac{1}{\sqrt{30}} & \frac{1}{\sqrt{30}} & -\sqrt{%
\frac{2}{15}} & -\sqrt{\frac{2}{15}} & -\sqrt{\frac{2}{15}} \\
0 & 0 & 0 & \frac{1}{\sqrt{6}} & -\sqrt{\frac{2}{3}} & \frac{1}{\sqrt{6}} \\
0 & -\sqrt{\frac{2}{5}} & \sqrt{\frac{2}{5}} & \frac{1}{\sqrt{10}} & 0 & -%
\frac{1}{\sqrt{10}} \\
\sqrt{\frac{3}{10}} & -\sqrt{\frac{3}{10}} & -\sqrt{\frac{3}{10}} & \frac{1}{%
\sqrt{30}} & \frac{1}{\sqrt{30}} & \frac{1}{\sqrt{30}}%
\end{array}}
\right)
\end{eqnarray}
For $(2,0)$, the highest weight projector is $\vert (2,0)(20) \rangle\langle
(2,0)(20) \vert$; Eqn.(\ref{Fforanys}) this time gives
\begin{equation}
F^{(s)}_0=\frac{\left(\sqrt{6}\right)^{s}}{6^{(s+1)/2}}\, ,\quad F^{(s)}_1=%
\frac{2^{-s}\left(\sqrt{\textstyle\frac{15}{2}}\right)^{s}}{(6\cdot
8)^{(s+1)/2}}\, ,\quad F^{(s)}_2=\frac{\left(\sqrt{\textstyle\frac{10}{3}}%
\right)^{s}} {(6\cdot 27)^{(s+1)/2}}\, .
\end{equation}

The expansion of $\hbox{\rm e}^{i\omega _{1}\hat{h}_{1}}$ contains three
terms:
\begin{eqnarray}
\hbox{\rm e}^{i\omega \hat{h}_{1}}&=&\textstyle\frac{1}{\sqrt{6}}\left(\hbox{\rm e}^{4i\omega }+2\hbox{\rm e}^{i\omega }
+3\e^{-2i\omega }\right) \hat{T}_{(0,0);\smb{0}0}^{(2,0)}
+\textstyle\sqrt{\frac{2}{15}}\,
\left( 2\hbox{\rm e}^{4i\omega }+\hbox{\rm e}^{i\omega }-3\e^{-i\omega }\right)
\hat T_{(1,1);\smb{0}0}^{(2,0)} \nonumber \\
&&\quad + \textstyle\sqrt{\frac{3}{10}}\,\left( \hbox{\rm e}^{4i\omega }-2\hbox{\rm e}%
^{i\omega }-3\e^{-2i\omega }\right) T_{(2,2);\smb{0}0}^{(2,0)}.
\end{eqnarray}%
The overlap matrix is found to be
\begin{equation}
g_{\sigma \sigma ^{\prime }}=\left(
\begin{array}{ccc}
\frac{14\pi }{3} & -\frac{2\sqrt{5}\pi }{3} & 0 \\
-\frac{2\sqrt{5}\pi }{3} & \frac{56\pi }{15} & -\frac{6\pi }{5} \\
0 & -\frac{6\pi }{5} & \frac{18\pi }{5}%
\end{array}%
\right) \,,
\end{equation}

The expansion coefficients for $f^{(s)}$ are then
\begin{equation}
\left(%
\begin{array}{c}
c^{(s)}_0 \\
c^{(s)}_1 \\
c^{(s)}_2%
\end{array}%
\right)= \left(%
\begin{array}{ccc}
\frac{1}{4\pi} & \frac{1}{4\sqrt{5}\pi} & \frac{1}{12\sqrt{5}\pi} \\
\frac{1}{4\sqrt{5}\pi} & \frac{7}{20\pi} & \frac{7}{60\pi} \\
\frac{1}{12\sqrt{5}\pi} & \frac{7}{60\pi} & \frac{19}{60\pi}%
\end{array}%
\right) \left(%
\begin{array}{c}
F^{(s)}_0 \\
F^{(s)}_1 \\
F^{(s)}_2%
\end{array}%
\right)
\end{equation}

For the special cases of $s=-1,0,1$, we find:
\begin{equation}
{\renewcommand{\arraystretch}{1.75}
\begin{array}{c|c|c|c}
& s=-1 & s=0 & s=1 \\ \hline
c^{(s)}_0= & \frac{1}{2 \sqrt{6} \pi } & \frac{90 \sqrt{6}+2 \sqrt{10}+9
\sqrt{15}}{2160 \pi } & \frac{8051}{31104\sqrt{6}\pi} \\ \hline
c^{(s)}_1= & \frac{\sqrt{2}}{\sqrt{15}\pi } & \frac{14 \sqrt{2}+63 \sqrt{3}%
+18 \sqrt{30}}{2160 \pi } & \frac{9701}{31104 \sqrt{30} \pi } \\ \hline
c^{(s)}_2= & \frac{1}{2 \pi }\sqrt{\frac{3}{10}} & \frac{38 \sqrt{2}+21
\sqrt{3}+6 \sqrt{30}}{2160 \pi } & \frac{3767}{31104 \sqrt{30} \pi }%
\end{array}
}\, .
\end{equation}

\section{Discussion: the general case}

We now briefly discuss the case of general irreps through the example of $\mathfrak{SU}(3)$
irreps of the type $(\lambda,\mu)$.  In such cases, the highest weight projector is invariant
under $\mathfrak{U}(1)\otimes \mathfrak{U}(1)$ transformations: the coset space is the
six--dimensional space $\mathfrak{M}=\mathfrak{SU}(3)/[\mathfrak{U}(1)\otimes \mathfrak{U}(1)]$.

The situation here is fundamentally different from the $(\lambda,0)$ case as the decomposition of $(\lambda,\mu)\otimes (\mu,\lambda)$ will contain some irreps more than once
\cite{su3tensordecomposition}.  These multiple copies of irreps are completely identical, with the results that two distinct operators may be mapped to the same phase space symbol and it becomes impossible to unambiguously reconstruct an operator from its symbol.

For instance, suppose we are working in a Hilbert space that carries the irrep $(1,1)$ of
$\mathfrak{SU}(3)$.  One can verify that
\beq
(1,1)\otimes (1,1)=(2,2)\oplus (3,0)\oplus (0,3)\oplus (1,1)\oplus (1,1)\oplus (0,0)\, ;
\eeq
the irrep $(1,1)$ occurs twice in the decomposition.

The highest weight projector is written in terms of tensors as
\beqa
&&\ket{(1,1)11;\hf}\bra{(1,1)11;\hf} \nonumber \\
&&\quad =\textstyle\frac{1}{\sqrt{8}}\,\hat T_{(0,0)\smb{0}0}^{(1,1)}
-\textstyle\frac{1}{2\sqrt{3}}\,\hat T_{(3,0);\smb{0}1}^{(1,1)}
-\textstyle\frac{1}{2\sqrt{3}}\,\hat T_{(0,3);\smb{0}1}^{(1,1)}
+\textstyle\sqrt{\frac{3}{14}}\,
\hat T_{(1,1)_1;\smb{0}1}^{(1,1)}-\textstyle\frac{1}{\sqrt{14}}\,\hat T_{(1,1)_1;\smb{0}0}^{(1,1)}\nonumber \\
&&\qquad +\textstyle\sqrt{\frac{2}{105}}\,\hat T_{(1,1)_2\smb{0}1}^{(1,1)}
+2\textstyle\sqrt{\frac{2}{35}}\,\hat T_{(1,1)_2;\smb{0}0}^{(1,1)}
+\textstyle\frac{1}{\sqrt{10}}\,\hat T_{(2,2);\smb{0}1}^{(1,1)}
-\textstyle\frac{1}{2}\sqrt{\frac{3}{10}}\,\hat T_{(2,2)\smb{0}0}^{(1,1)}\, .
\eeqa
The tensors $\hat T_{(1,1)_1\smb{0}1}^{(1,1)}$ and $\hat T_{(1,1)_2;\smb{0}1}^{(1,1)}$
transform identically under $\mathfrak{SU}(3)$, as do
the tensors $\hat T_{(1,1)_1;\smb{0}0}^{(1,1)}$ and $\hat T_{(1,1)_2;\smb{0}0}^{(1,1)}$.  There is no choice of basis in the decomposition that will make one copy of $(1,1)$ disappear.  Thus, tensors like
$\hat T_{(1,1)_1;\smb{0}1}^{(1,1)}$ and $\hat T_{(1,1)_2;\smb{0}1}^{(1,1)}$, even through they are orthogonal,
will be mapped to identical phase space symbols.  In particular, the
$Q$-symbols
$Q_{\hat T_{(1,1)_1;\ssmb{0}1}^{(1,1)}}=\bra{(1,1)11;\hf}\Lambda^\dagger (\Omega)\,\hat T_{(1,1)_1;\smb{0}1}^{(1,1)}\, \Lambda(\Omega)\ket{(1,1)11;\hf}$ will be
proportional to $Q_{\hat T_{(1,1)_2;\ssmb{0}1}^{(1,1)}}$.

\section{Conclusion}

This paper presents an algorithm to find a kernel $w^{(s)}(\Omega)$ that allows one-to-one mapping between operators in the Hilbert space carrying an irrep $(\lambda,0,\ldots,0)$ of $\mathfrak{SU}(n)$ and c-functions defined on the corresponding phase space $\mathfrak{SU}(n)/\mathfrak{U}(n-1)$.

For irreps of the $(\lambda,0,\ldots,0)$ type, our classical manifold is the "canonical" one identified from properties of the coherent states, as suggested by \cite{Onofri}. For the more general representations the mapping no such mapping to the canonical phase space is possible and different Wigner--like distributions should be constructed (as for instance in {\cite{mukunda}\cite{mukunda2}).

Beyond the obvious computationally attractive extension of the $P$--, $Q$-- and Wigner functions and their possible applications to describe collection of indistinguishable $n$-level atoms and BEC \cite{Viscondi}, our approach could also be the pathway toward understanding the asymptotic forms and contraction limits of these objects, much as was done in \cite{Sergei}. Our prescription can also be the starting point to develop the $\star$--product for $\mathfrak{SU}(3)$.

We would like to thank Lakehead University for partial support of this work, and Mr. Zach Medendorp for his
help with various computational aspects of this work.
HdG is supported by NSERC of Canada.  ABK thanks CONACYT Mexico for support through grant \#106525.


\bigskip

\begin{thebibliography}{99}

\bibitem{molecule} A. V. Gorshkov \emph{et al.}, Nature Physics \textbf{6} (2010) 289,

\bibitem{Viscondi} T. F. Viscondi, K. Furuya and M. C. de Oliveira, Eur. Phys. Lett \textbf{90} (2010) 10014,

\bibitem{Wigner} E. P. Wigner Phys. Rev. \textbf{40}(1932) 749,

\bibitem{reviews} M. Hillery, R. F. O'Connell, M. O. Scully and E. P. Wigner,
Phys. Rep. \textbf{106} (1984) 121,

\bibitem{Lee} H.-W. Lee, Phys. Rep. \textbf{259} (1995) 147,

\bibitem{Perelomov} A. Perelomov, \textit{Generalized Coherent states and
their applications} (Springer-Verlag Berlin, 1986),

\bibitem{gilmore} F.T. Arecchi, E.Courtens, R. Gilmore and H. Thomas, Phys.
Rev. A \textbf{6} (1972) 2211,

\bibitem{haake} F. Haake and R. J. Glauber, Phys. Rev. A \textbf{5} (1972) 1457;
\emph{ibid},Phys. Rev. A \textbf{13} (1976) 357,

\bibitem{Varilly} J. C. V\'{a}rilly and J. M. Gracia-Bond{\'{\i}}a, Ann.
Phys. (N.Y.) \textbf{190} (1989) 107,

\bibitem{Weigert} S. Heiss and S. Weigert, Phys. Rev. A \textbf{63} (2000) 012105,

\bibitem{Agarwal} G. S. Agarwal, Phys. Rev. A \textbf{24} (1981) 2889,

\bibitem{Wolf} K. B. Wolf, Opt. Commun. \textbf{132}(1996) 343; N. M. Atakishiyev,
S. M. Chumakov and K. B. Wolf, J. Math. Phys. \textbf{39} (1998) 6247,

\bibitem{Brif} C. Brif and A. Mann, Phys. Rev. A \textbf{59} (1999) 971,

\bibitem{mukunda} N. Mukunda, G. Marmo, A. Zampini, S. Chaturvedi and R. Simon,
J.Math.Phys. \textbf{46} (2005) 012106,

\bibitem{mukunda2} S. Chaturvedi, E. Ercolessi, G. Marmo, G. Morandi, N. Mukunda,
and R. Simon, J.Phys.A \textbf{39} (2006) 1405,

\bibitem{Moyal} J. E. Moyal, Proc. Cambridge Philos. Soc. \textbf{45} (1949) 99,

\bibitem{Stratonovich} R. L. Stratonovich, Sov.\ Phys.\ JETP \textbf{31} (1956)
1012,

\bibitem{Beresin} P. A. Beresin, Comm. Math. Phys. \textbf{40} (1975) 153,

\bibitem{SU3harmonic} Douglas Francis Holland, J.~Math.~Phys. \textbf{10} (1969) 531; Mirza A. B. B\'eg
and Henri Ruegg, J.~Math.~Phys. \textbf{6} (1965) 677,

\bibitem{su3Wigner} D. J. Rowe, B. C. Sanders and H. de Guise, J.~Math.~Phys. \textbf{40} (1999) 3604,

\bibitem{Onofri} Enrico Onofri,  J.~Math.~Phys. \textbf{16} (1975), 1087,

\bibitem{AlfredoLuis} Alfredo Luis, J. Phys. A: Math. Theor. \textbf{41} (2008) 495302,

\bibitem{su3tensordecomposition} Maria S M Wessl\'en, J. Phys.: Conf. Ser. \textbf{175} (2009) 012015;
D. Speiser, \emph{Group Theoretical Concepts and Methods in Elementary Particle Physics}, Lectures of the
Istanbul Summer School of Theoretical Physics ed F G\"{u}rsey (New York: Gordon and Breach, 1962) pp 201-76;
Michael F. O'Reilly, J.~Math.~ Phys. \textbf{23} (1982) 2022,

\bibitem{Sergei} A. B. Klimov and S. M. Chumakov, JOSA \textbf{A17} (2000) 2315.

\end{thebibliography}
\end{document}